\documentclass[structabstract]{aa} % 2 column
\usepackage{graphicx} 
\usepackage{float}

\usepackage{epstopdf}
\usepackage{txfonts}
\usepackage{array}
\usepackage{natbib}
\usepackage{units}
\usepackage{url}
\usepackage{bm}
\usepackage{lscape}
\usepackage{supertabular,booktabs}
\usepackage{comment}
\usepackage{multirow}
\usepackage{nccmath}
\usepackage{amsmath, amstext}
\usepackage{hyperref}
\usepackage{amssymb}
\usepackage{amsmath}
\usepackage{threeparttable}
\usepackage{float}
\usepackage{mathtools}
\usepackage{color}
\usepackage{cancel}
\usepackage{caption}
\begin{document}

\title{Is there a non-stationary $\gamma$-ray emission zone 42~pc away from the 3C~279 core?}

% List of authors
\author{V. M. Pati\~no-\'Alvarez\inst{1,2}\thanks{E-mail:victorp@inaoep.mx}
\and
S. A. Dzib\inst{1}
\and
A. Lobanov\inst{1,3}
\and
V. Chavushyan\inst{2}}
%\newauthor 
% List of institutions
\institute{$^1$Max-Planck-Institut f\"ur Radioastronomie, Auf dem H\"ugel 69,
 D-53121 Bonn, Germany
\\
$^2$Instituto Nacional de Astrof\'isica \'Optica y Electr\'onica (INAOE), Apartado Postal 51 y 216, 72000 Puebla, M\'exico
\\
$^3$Institut f\"ur Experimentalphysik, Universit\"at Hamburg, Luruper Chaussee 149, 22761 Hamburg, Germany
}
% \\
% \email{patinoavm@mpifr-bonn.mpg.de}

% These dates will be filled out by the publisher
%\date{Accepted XXX. Received YYY; in original form ZZZ}

% Enter the current year, for the copyright statements etc.
%\pubyear{2017}

% Don't change these lines
%\begin{document}
%\label{firstpage}
%\pagerange{\pageref{firstpage}--\pageref{lastpage}}
%\maketitle

%%%%%%%%%%%%%%%%%%%%%%%%%%%%%%%%%%%%%%
%%%%%%%%%%%%%%%% ABSTRACT %%%%%%%%%%%%%%%%
%%%%%%%%%%%%%%%%%%%%%%%%%%%%%%%%%%%%%%

\date{Received 2018; }
\abstract
{We investigate the relationship between the variable gamma-ray emission and jet properties in the blazar 3C~279, by combining the Fermi/LAT data spanning a period of eight years with concurrent radio measurements made at multiple epochs with VLBA at 15 and 43 GHz within the MOJAVE and VLBA-BU monitoring programs. The aim of this paper is to compare the flux variability of the different components found in the VLBA observations, to the variability in the gamma-rays. This analysis helps to investigate whether any of the jet components can be associated with the gamma-ray variability. Through Spearman rank correlation we found that the gamma-ray variability is correlated with a particular region (feature B, in the MOJAVE images) downstream from the observed base (core) of the jet. This jet component is therefore a likely location at which an important fraction of the variable gamma-ray emission is produced. We also calculated the average proper motion of the component with respect to the VLBA core and found that it moves at an apparent superluminal velocity of $(3.70 \pm 0.35)c$, implying that one of the gamma-ray emission zones is not stationary. This jet component is also found between 6.86 mas and 8.68 mas, which translates to a distance from the radio core of at least 42 pc.}

\keywords{galaxies: active --- galaxies: jets --- gamma-rays: galaxies --- quasars: individual (3C~279)}

\titlerunning{Gamma-VLBA Relationship in the FSRQ 3C~279}

\maketitle

%%%%%%%%%%%%%%%%%%%%%%%%%%%%%%%%%%%%%%
%%%%%%%%%%%%%% INTRODUCTION %%%%%%%%%%%%%%%
%%%%%%%%%%%%%%%%%%%%%%%%%%%%%%%%%%%%%%

\section{Introduction}

Blazars are a class of active galactic nuclei whose relativistic jets are pointing very close to the observer's line of sight \citep{Urry1995}. At a redshift of 0.5362, 3C~279 is among the most studied blazars, due to its high brightness, variability, and peculiar optical spectrum \citep[e.g.][]{Torrealba2012}. 3C~279 was also one of the first gamma-ray emitting quasars discovered with the Compton Gamma-Ray Observatory \citep{Hartman1992}. It has been known for decades, that this source is highly variable from radio frequencies \citep{Pauliny-Toth1966}, to gamma-rays \citep{Hartman1992}. For this reason, several intensive multiwavelength campaigns and theoretical studies have led to important revelations about the physical properties of 3C~279 {\citep[e.g.][and references therein]{Gurwell2007, Chatterjee2008, Marscher2008,  Lister2009, Lister2013, Smith2009, Abdo2010c, Richards2011, Bonning2012, Nolan2012, Bottcher2013, Patino-Alvarez2013a, Patino-Alvarez2018, Acero2015, Jorstad2017}.
 
3C~279 is a highly variable radio source, however, at radio wavelengths, the variability of this source appears to be more moderate than the corresponding variability in the gamma-ray and optical bands during the time period between 2008 and 2010 \citep{Hayashida2012}. The excess variance \citep[$F_{var}$,][]{Vaughan2003} in the radio regime is quite modest: $0.145$ at 37 GHz, $0.165$ at 15 GHZ, and $0.104$ at 5 GHz, in contrast to the gamma-rays $F_{var}=0.695$. \cite{Hayashida2015} and \cite{Patino-Alvarez2018} reported fast gamma-ray variability in 3C~279 during 2013-2014. Several other observational and theoretical studies have been carried out in order to determine both the nature and the stratification of the emission at different bands in 3C~279 (i.e. emission mechanisms and origin of the radiation) \citep[e.g.][]{Lindfors2006,Larionov2008,Collmar2010,Aleksic2011,Janiak2012,Hayashida2012,Hayashida2015,Rani2018}. These studies point to the possible presence of multiple emission regions, however, despite these studies, a general consensus about the location of the gamma-ray production zone in 3C~279 does not yet exist.

Previous studies have attempted to link the variability in the optical and gamma-ray continuum emission to kinematic changes in VLBI (Very Long Baseline Interferometry) radio images, in several individual objects, including 3C~390.3 \citep{Arshakian2010}, 3C~120 \citep{Leon-Tavares2010}, and 3C~345 \citep{Schinzel2012}. These papers showed that the passage of a moving jet component through a stationary shock can trigger flares in emission bands other than the radio. 

Thanks to extensive multi-epoch surveys like the 15 GHz MOJAVE (Monitoring Of Jets in Active galactic nuclei with VLBA\footnote{Very Long Baseline Array of the National Radio Astronomy Observatory, Socorro NM, USA} Experiments, see \citealp{Lister2009,Lister2013}) program; and the 43 GHz Boston University VLBA-Blazar monitoring program \citep[VLBA-BU, see][]{Jorstad2017} the community has at its disposal a very rich database of VLBA images and reliable measurements of jet proper motions for a number of sources, which allow researching intricate structural changes in parsec-scale jets, e.g. transverse motions, jet acceleration, and other morphological changes.

High angular resolution VLBA observations at 15 and 43 GHz, have shown that 3C~279 is a core-dominated blazar, with a complex multi-component jet structure \citep[see Figures~\ref{fig:mojave} and \ref{fig:BU}, and also][]{Lister2009,Jorstad2017}. In order to investigate whether the distinct emitting regions (jet components) observed in the VLBA maps are related to the gamma-ray emission production, we study the flux variability of these components and compare it to the variability observed in the gamma-rays.

%%%%%%%%%%%%%%%%%%%%%%%%%%%%%%%%%%%%%%
%%%%%%%%%%%%%% OBSERVATIONS %%%%%%%%%%%%%%%
%%%%%%%%%%%%%%%%%%%%%%%%%%%%%%%%%%%%%%

\section[]{Observations}

We used gamma-ray data from the Fermi Large Area Telescope \citep[Fermi-LAT,][]{Abdo2009}, the VLBA data from the MOJAVE program \citep{Lister2009}, data from the VLBA-BU program \citep{Jorstad2017}, as well as data from the OVRO monitoring program \citep{Richards2011}. These databases are publicly available. Hereafter, all the dates we present are in the format $\rm{JD}_{245}=
\rm{JD}-2'450'000$.

\subsection{Gamma-Rays}
\label{gamma}

The gamma-ray light curve in the energy range from 0.1 to 300 GeV was built by using data from the Fermi-LAT. The Fermi-LAT data were reduced and analyzed with the Fermitools version 1.0.2. The region of interest was selected to be $15^{\circ}$ radius, centered at the position of 3C~279. The minimization was 
done through a maximum likelihood algorithm, and the source spectrum was modeled as a log-parabola, since it fits the spectrum of 3C 279 better than a power-law model, and is also the spectrum type listed in the 4FGL catalogue \citep{4FGL}. We included all sources within $15^{\circ}$ of 3C~279, 
extracted from the 4FGL catalogue \citep{4FGL}, with their normalization kept free, while sources within $5^{\circ}$ had both, their normalizations and spectral indices kept free. We applied the currently recommended set of 
the instrument response functions along with the latest diffuse and isotropic background model files. For demonstrations purposes only, the light curve in Figure~\ref{curves} was built with a time bin of 7 days, in order to increase the S/N ratio and kept only the fluxes from time bins with a TS $>$ 25 (detection at 5$\sigma$ or higher). Upper limits were not calculated for the bins with TS $<$ 25, because they are not suitable for correlation analysis. The light curve comprises the time-range since the launch of Fermi in June 2008 up to February 2017. This light curve is shown in Figure~\ref{curves} panel~(a).

\subsection{VLBA 15 GHz Observations}
\label{15GHz_data}

The MOJAVE VLBA observations were carried out in 24 hour-long segments at monthly intervals, and in each segment, a small subset of the total AGN sample is observed \citep{Lister2009,Lister2016}. The subsets were chosen on the basis of the individual rates of angular jet motion, with several very fast ($>2$ mas/year) jets being observed once per month, while the slowest jets are observed only once every few months, as is the case for 3C~279. This resulted in a total of 21 observations of 3C~279 in the time period of the gamma-ray light curve, as described in Section~\ref{gamma}. The natural weighted and non-tapered FITS images were obtained from the MOJAVE website\footnote{\url{http://www.physics.purdue.edu/astro/MOJAVE/index.html}}. 

The maps used in this work show that there are two main emission features (see Figure~\ref{fig:mojave}); we designate the brightest as feature A, and the secondary as feature B. Since feature A is where the radio core is expected to be, we model this feature by fitting 2-D Gaussian components using the CASA software \citep{McMullin2007}, in order to single out the flux of the radio core at 15 GHz. Modeling of feature B was more difficult due to its considerably lower brightness. We were also unable to find consistency in our feature B Gaussian model fits between epochs. Therefore, we carried out our analysis using an integrated value of the flux from feature B. The light curves for the core and the feature B components of the MOJAVE maps are shown in Figure~\ref{curves} panels (d) and (e), respectively. For comparison, we also show in Figure~\ref{curves} panel (f), the light curve of the total flux density at 15 GHz from the Owens Valley Radio Observatory (OVRO) blazar monitoring program \citep{Richards2011}.

\begin{figure*}
\begin{center}
\includegraphics[width=0.98\textwidth]{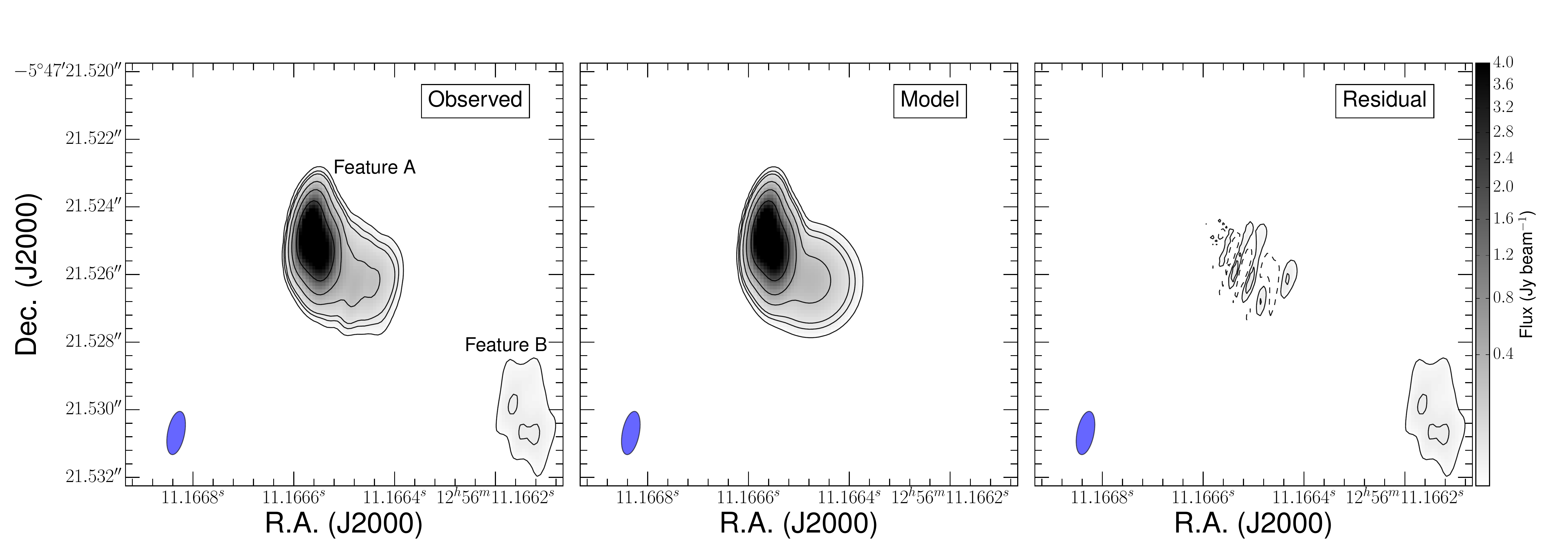}
\caption{Left panel: 15 GHz VLBA intensity map from the MOJAVE observation of July 16, 2016. Middle panel: Model of Feature A using 2D-Gaussian components. Right Panel: Residual image, where Feature B is visible in the bottom right corner of the map.}
\label{fig:mojave}
\end{center}
\end{figure*}

\subsection{VLBA 43 GHz Observations}

The ongoing VLBA-BU program comprises monthly VLBA observations at 43 GHz. 
The observations, commenced in June 2007, typically span 24 hours during which roughly 
33 AGN are observed, with about 45 minutes spent on each target. There are 84 observations of 
3C~279 within the time-range of the gamma-ray light curve. The details on 
this program, the observations, and data reduction, are described in \cite{Jorstad2017}. The FITS images
were obtained from the VLBA-BU program website\footnote{\url{https://www.bu.edu/blazars/VLBAproject.html}}.
These were also analyzed using the CASA software.

As is illustrated in Figure~\ref{fig:BU}, the total emission in the 43 GHz maps does not come from a single component. In order to characterize the 43 GHz jet structure we used the CASA task IMFIT to model the radio emission using 2-D Gaussians. We were able to single-out the flux of the core in each map, thus creating a light curve of the core flux, which is shown along with the total flux in Figure~\ref{curves} panel (b).

Up to mid-2012, the flux density of the core clearly dominated the total emission in the 43 GHz maps (see Figure \ref{curves}). 
An ejection of a new jet component is clearly visible in 2013 \citep[see Figure~\ref{fig:BU}, and][]{Jorstad2017}. During the initial stages this component became brighter than the core, and dominated the total flux density. This component can be tracked over 17 consecutive observations while gradually becoming fainter until it is no 
longer observable. This ejection is labeled as C31 in \cite{Jorstad2017}, and hereafter we refer to this ejection as such. The light curve for the flux density of C31 is shown in Figure~\ref{curves} panel (c). 
Because of the limited resolution of the 43 GHz VLBA maps (with a beam size of a few parsecs), it is not possible to assess with these observations, 
if the ejection of C31 actually occurred during mid-2012, or if the core flux measured before was actually a combination of both, the core 
and the new jet component.

\begin{figure*}
\begin{center}
\includegraphics[width=0.98\textwidth]{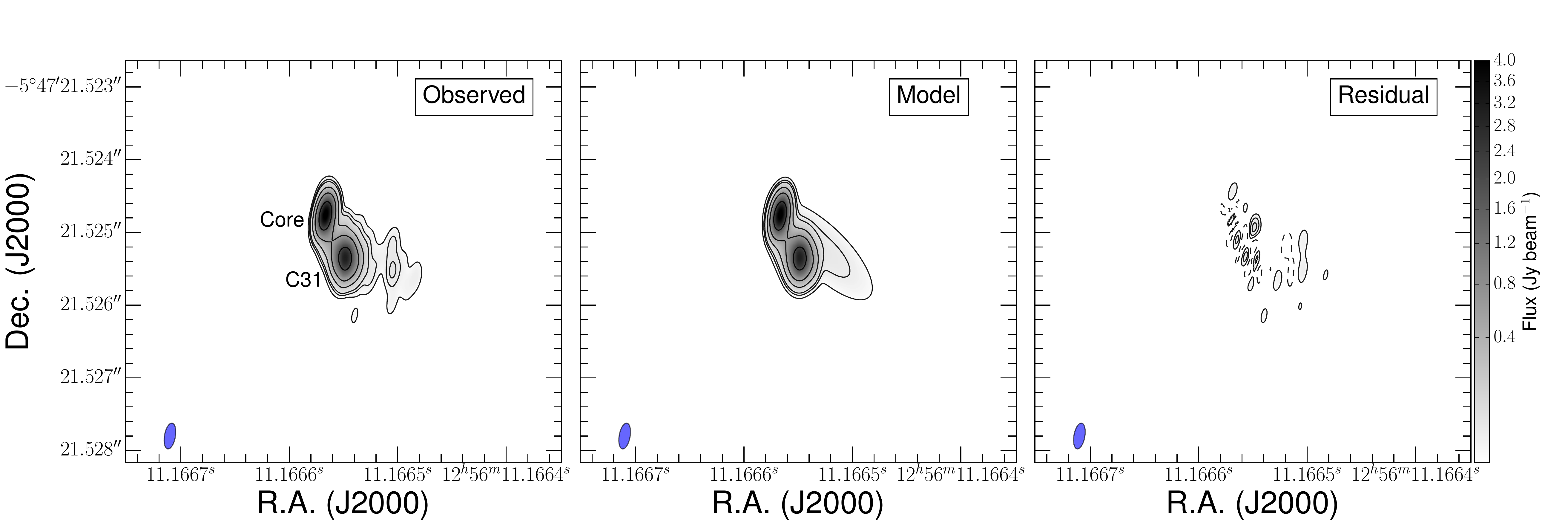}
\caption{Left Panel: 43 GHz VLBA intensity map of 3C~279 from the BU observation of August 26, 2013. Middle Panel: Model of the observed emission using 2-D Gaussian components. Right Panel: Residual image.}
\label{fig:BU}
\end{center}
\end{figure*}

\subsection{OVRO 15 GHz Observations}

We obtained fully reduced data from the OVRO blazar monitoring program\footnote{\url{http://www.astro.caltech.edu/ovroblazars/data/data.php}}. Full details of the data reduction and instrument can be found in \cite{Richards2011}. The light curve can be seen in panel (f) of Figure~\ref{curves}.

\begin{figure*}[h!tp]
\begin{center}
\includegraphics[width=0.95\textwidth]{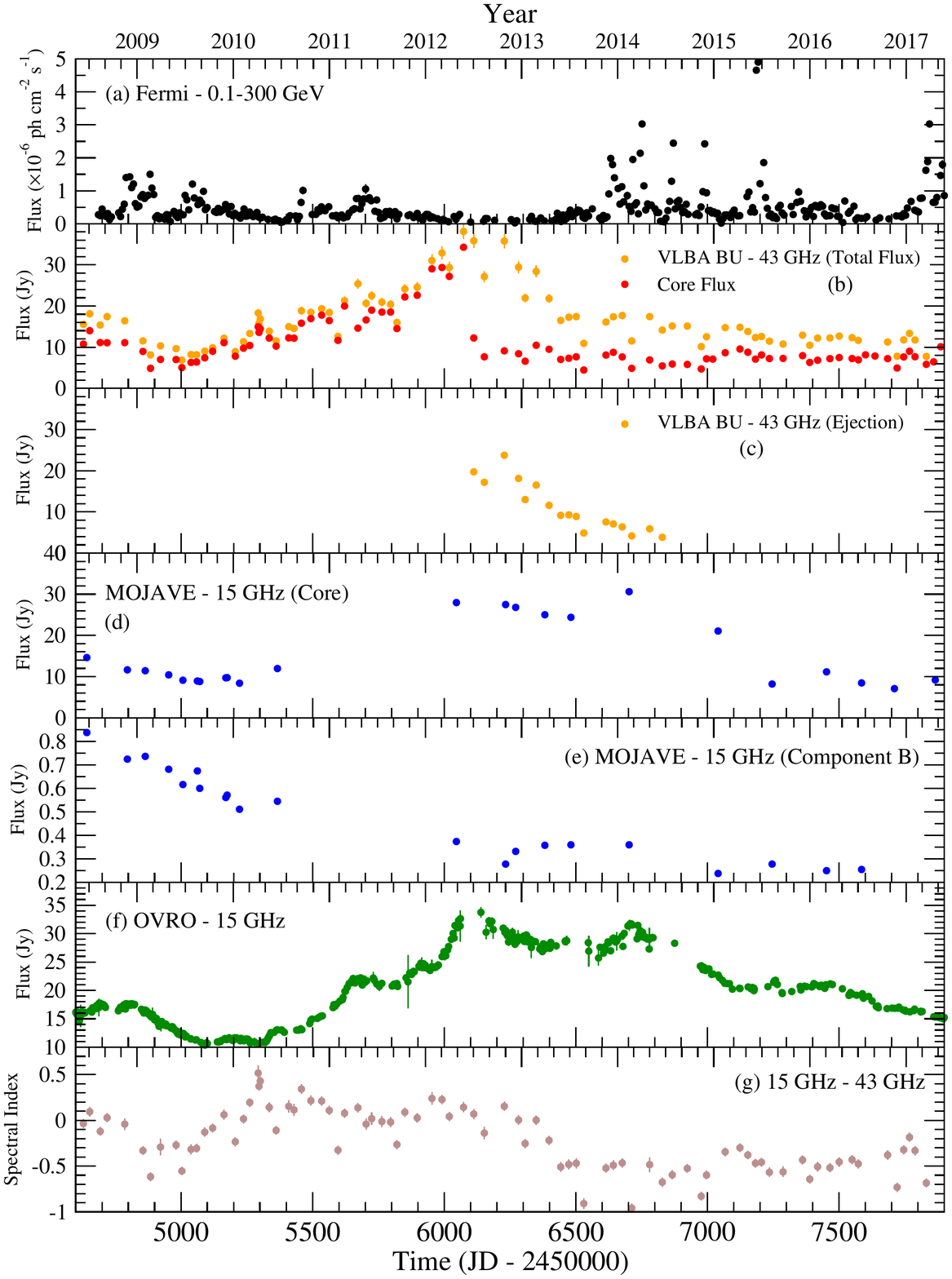}
\caption{Multiwavelength light curves of 3C~279. The band of observation and the origin of the data is labeled inside each panel.}
\label{curves}
\end{center}
\end{figure*}

%%%%%%%%%%%%%%%%%%%%%%%%%%%%%%%%%%%%%%
%%%%%%%%%%%% CORRELATION %%%%%%%%%%%%%%%%%
%%%%%%%%%%%%%%%%%%%%%%%%%%%%%%%%%%%%%%

\section{Correlation Analysis}
\label{CorrSec}

In order to understand if any of the features observed in the MOJAVE and BU maps can be directly related to the production of gamma-ray emission, we perform Spearman correlation tests between the simultaneous flux densities of each component with the gamma-ray fluxes. In order to do this, we calculated gamma-ray light fluxes in bins centered at the observation times of the BU and MOJAVE data. We tested incremental bin sizes, starting with the small variability time-scales reported in \cite{Ackermann2016} and \cite{Pittori2018}; always maintaining the center as the observation date of the VLBA observations, until the TS for each respective bin was higher than 25.

The relations between the 15 GHz flux densities of the core and feature B of the MOJAVE maps and the calculated simultaneous gamma-ray fluxes are shown in the top panels of Figure~\ref{correlation}. The respective relations between the fluxes of the core and the component C31 in the BU maps, with the simultaneous gamma-ray fluxes are shown in the bottom panels of Figure~\ref{correlation}. The results of the Spearman correlation tests are summarized in Table~\ref{Spearman}.

\begin{figure*}[h!]
\begin{center}
\includegraphics[width=0.9\textwidth]{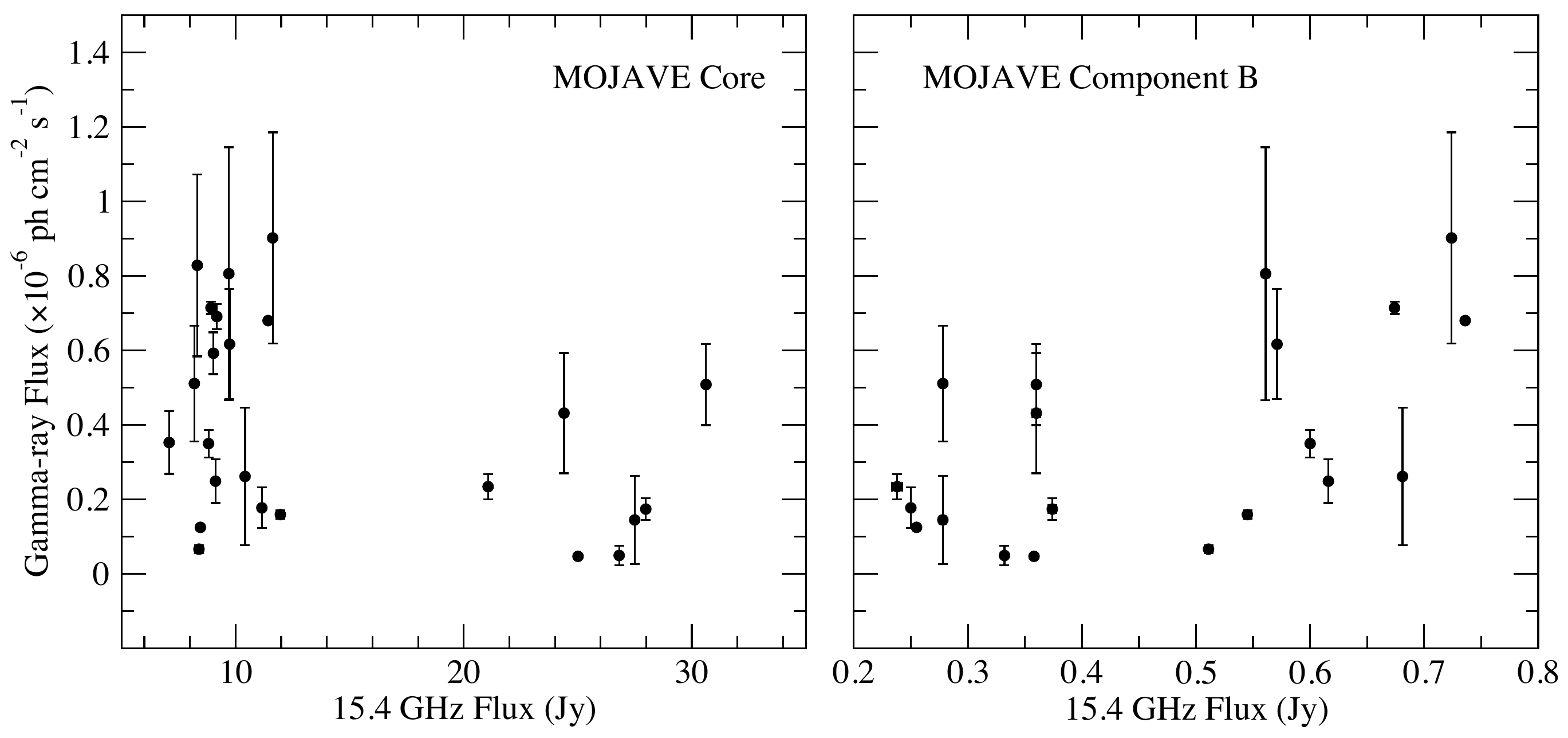}
\includegraphics[width=0.9\textwidth]{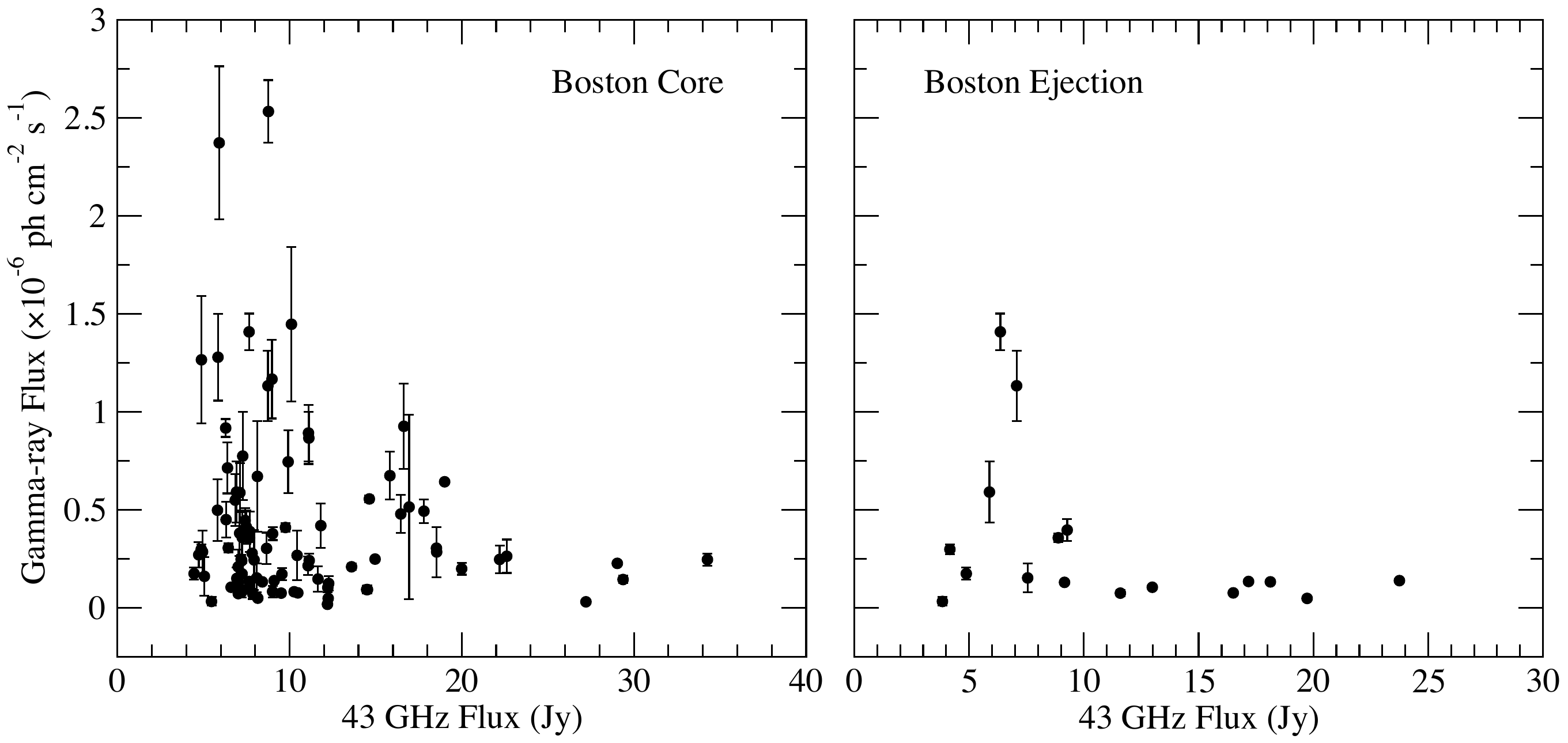}
\caption{Top left panel: Gamma-ray flux vs. radio flux for the core in the MOJAVE maps. Top right panel: Gamma-ray flux vs. radio flux for component B in the MOJAVE maps. Bottom left panel: Gamma-ray flux vs. radio flux for the core in the BU-VLBA maps. Bottom right panel: Gamma-ray flux vs. radio flux for C31 in the BU-VLBA maps.}
\label{correlation}
\end{center}
\end{figure*}

We consider a given correlation to be significant if the probability (p-value) of obtaining a specific correlation coefficient by chance is lower than 5\% ($p<0.05$). Under this criterion, only the feature B in the MOJAVE maps yields significant correlations (see Table~\ref{Spearman}). We find a strong and significant correlation between the gamma-ray emission and feature B of the MOJAVE maps, with a correlation coefficient of 0.57. This result can be interpreted as indicating that about 33\% of the variability observed in the gamma-rays is driven by the variability in the flux density of feature B. In order to discard the possibility that the significant correlations obtained are driven by individual data-points we performed a jacknife n-1 analysis, i.e., repeat the correlation as many times as data-points available, but discarding a different data-point each time. Our results show that, in all cases, the correlation holds significance at a p-value lower than 3\%, with correlation coefficients between 0.51 and 0.65, implying strong correlations in all cases. This allows us to confirm that the variability in the gamma-rays is related to feature B of the MOJAVE maps. Based of this result, and the fact that the dominant gamma-ray emission process in this source is found to be Inverse Compton Scattering \citep{Patino-Alvarez2018} we conclude that a gamma-ray emission site is most likely contained within Feature B, which is located downstream of the radio core.

\begin{table}
\centering
\caption{Spearman correlation coefficients obtained between the fluxes of the features in the VLBA maps and the simultaneous gamma-ray fluxes}.
\begin{tabular}{|c|c|c|c|}
\hline
\textbf{VLBA Feature} & \textbf{R} & \textbf{R}\bm{$^2$} & \textbf{p-value}  \\
\hline
MOJAVE Core & -0.308 & 0.095 & 0.143 \\
\hline
MOJAVE Feature B & 0.573 & 0.328 & 0.008 \\
\hline
BU Core & -0.128 & 0.016 & 0.231 \\
\hline
BU Ejection & -0.400 & 0.160 & 0.112 \\

\hline
\end{tabular}
\label{Spearman}
\end{table}

%%%%%%%%%%%%%%%%%%%%%%%%%%%%%%%%%%%%%%%%%%%%%%%%%%%%%%%%%%
%%%%%%%%%%%%%% SPECTRAL INDEX CHANGE IN THE RADIO EMISSION %%%%%%%%%%%%%%%
%%%%%%%%%%%%%%%%%%%%%%%%%%%%%%%%%%%%%%%%%%%%%%%%%%%%%%%%%%

\section{Spectral Index Change in the Radio Emission and the Multiwavelength Perspective}

The substantial flaring activity observed in 3C~279 during 2012 can be explained by an increase
in the Lorentz factor of the jet, which at the same time can cause an increase in the gamma-ray
opacity via electron-positron pair production \citep{Patino-Alvarez2017,Patino-Alvarez2018}. 
This matches well with the low emission levels observed in the gamma-ray light curve between 2012 and 2013 (see Figure~\ref{curves} panel~a).

In order to test if our hypothesis is true, we combined the total VLBI flux densities at 43 GHz and the 15 GHz OVRO flux densities to estimate the spectral index. Since the 15 GHz light curve is very well sampled, it was interpolated to the epochs of the 43 GHz observations. The resulting 15-43 GHz spectral indices, defined as $S_{\nu} \propto \nu^{\alpha}$, are shown in Figure~\ref{curves} panel~(g).

As can be seen in the panels (b) and (f) of Figure~\ref{curves}, the variability of the total VLBI flux density at 43 GHz and the 15 GHz emission (OVRO), match each other very well up to the end of 2012. Afterwards, there is a clear decrease in the 43 GHz jet emission (Figure~\ref{curves} - panel~b), while the 15 GHz emission continues to remain active and increases starting around $\rm{JD}_{245}=6600$. This causes a decrease in the spectral index around the same epoch.

This change in the spectral index can be a result of a shift in the energy distribution of the electron population to lower energies, which is an expected consequence of the scenario mentioned above. We point out that the spectral index started a decreasing trend approximately at the same time than the onset of decreasing 43 GHz emission ($\sim\rm{JD}_{245}=6100-6200$), which supports this scenario.

%%%%%%%%%%%%%%%%%%%%%%%%%%%%%%%%%%%%%%%%%%%%%%%%%%%%%%%%%%
%%%%%%%%%%%%% RELATIVE MOVEMENT BETWEEN RADIO COMPONENTS %%%%%%%%%%%%%%
%%%%%%%%%%%%%%%%%%%%%%%%%%%%%%%%%%%%%%%%%%%%%%%%%%%%%%%%%%

\section{Relative Motion between MOJAVE Radio Components}

Since the jet feature B is likely to be a gamma-ray emission region, exploring its kinematics can yield further insights on its nature. We cannot find a prior analysis of its kinematics in the literature. Therefore, we use the VLBA observations from MOJAVE for estimating the proper motion of the feature B. It has been demonstrated that this kind of relative positional measurements provide very accurate results \citep[e.g.,][]{Rodriguez2017}. To obtain the relative proper motion of the feature B, first, we determine the position of the brightest spot in Feature B using the task IMFIT in CASA. We then subtracted the position of the core from that of Feature B at each epoch. Finally, we perform linear fits to the resulting core offset of the feature B in R.A. and in Dec. in order to estimate the change in position of Feature B as a function of time. The measured offsets and the line fits are shown in Figure~\ref{PM}.

The measured proper motion for Feature B in R.A. is: $-130.2 \pm 12.7 \; \mu \rm{as/year}$, and in Dec. $-130.8 \pm 12.7 \; \mu \rm{as/year}$. By applying a cosmology with $\rm{H_0}=73.0 \; km/s/Mpc$, $\Omega_m=0.27$, and $\Omega_{\Lambda}=0.73$, these proper motions corresponds to a total apparent velocity of $(3.70 \pm 0.35)c$. This result implies that the gamma-ray emission site(s) contained within Feature B are moving relativistically, which is an important detail when trying to model the gamma-ray emission.

\begin{figure*}
\begin{center}
\includegraphics[width=0.9\textwidth]{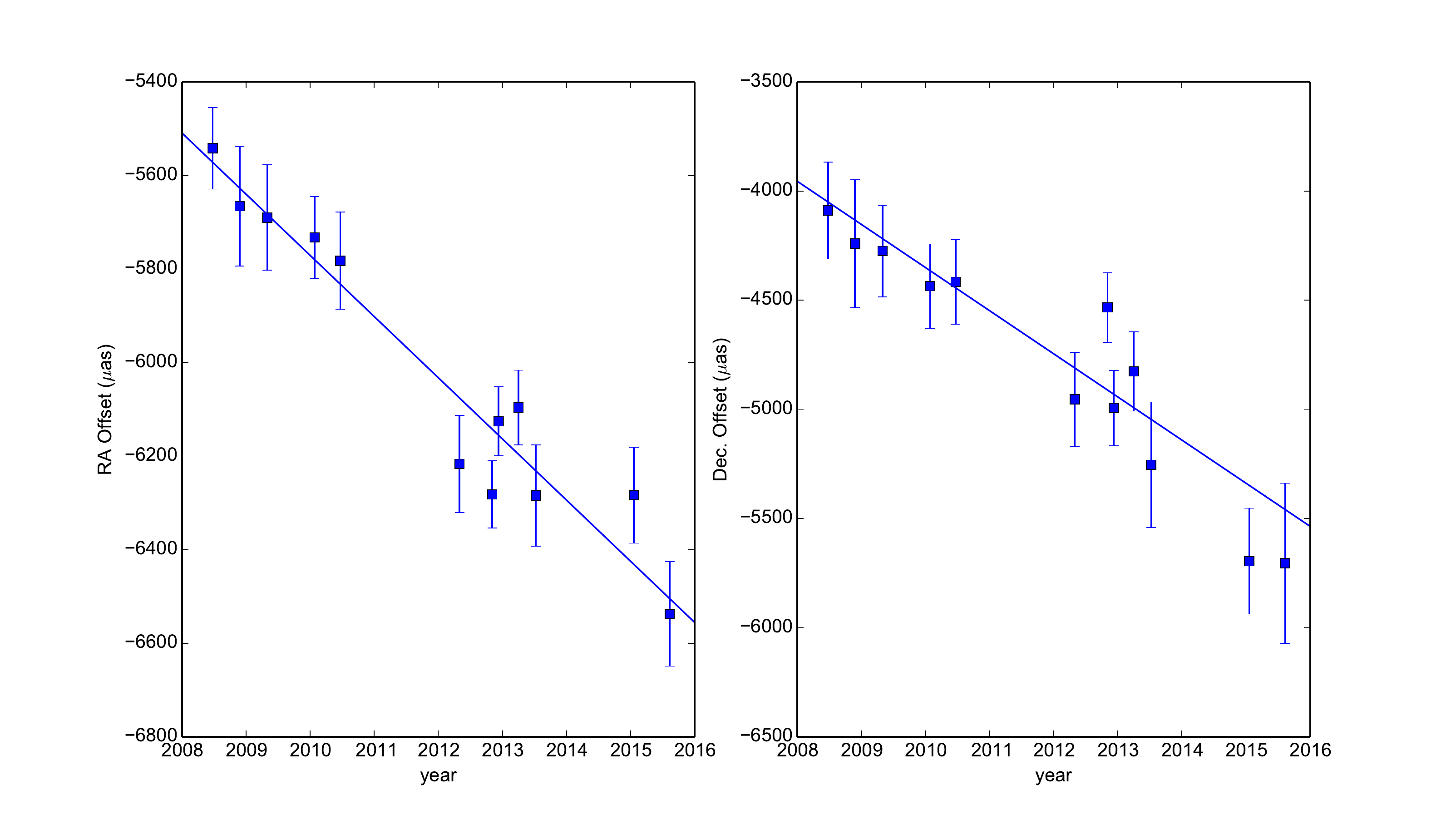}
\captionsetup{justification=centering}
\caption{Left panel: Separation of Feature B from the core, in right ascension, at the different observation epochs. Right panel: Separation of Feature B from the core, in declination, at the different observation epochs. The line represents a chi-square linear fit to the separations, from which we are able to obtain the velocity of Feature B.}
\label{PM}
\end{center}
\end{figure*}

%%%%%%%%%%%%%%%%%%%%%%%%%%%%%%%%%%%%%%
%%%%%%%%%%% RESULTS AND DISCUSSION %%%%%%%%%%%%
%%%%%%%%%%%%%%%%%%%%%%%%%%%%%%%%%%%%%%

\section{Conclusions}
\label{Conclusions}

We have combined Fermi-LAT measurements of the gamma-ray flux in 3C~279 with VLBI images of parsec-scale structure of the jet  and used this combination to identify the likely region responsible for the variable gamma-ray emission in this blazar. We analyzed the VLBA data from the MOJAVE survey and the Boston University (BU) monitoring program. We identified two prominent emission features in the MOJAVE maps. Feature A was modeled and decomposed in order to isolate the core flux. We also modeled the 43 GHz BU maps in order to obtain the core flux, as well as any distinct emitting regions inside the jet downstream from the core. We identified a component ejected around 2012 \citep[labelled C31 in the analysis of][]{Jorstad2017} and tracked its evolution over 17 consecutive epochs.

We compare the flux variability of feature B and the core of the MOJAVE maps, as well as the core and component C31 of the BU maps, to that of the gamma-ray emission via a Spearman rank correlation test. Our results imply that the feature B is likely related to the variability in the gamma-ray emission. Based on this we suggest that an emission zone responsible for part of the gamma-ray variability is contained within feature B of the MOJAVE maps. This is the first time that a long-lasting gamma-ray emission region has been clearly identified in a pc-scale map of a blazar.

Also important, is that the radio component that we propose may contain a gamma-ray emission zone, is well beyond the central parsec. The measured distances to the MOJAVE feature B go between 6.86 mas and 8.68 mas, which translates roughly to distances between 42 and 54 pc, depending on the cosmology applied.

We find evidence of a change in the spectral index of the radio emission, after a period of low gamma-ray emission. This change of spectral index is expected in the scenario of a shift in the energy distribution of the electron population, to lower energies, due to an increase in the cooling via synchrotron emission. This further supports the hypothesis by \citet{Patino-Alvarez2017}, and \citet{Patino-Alvarez2018}, that an increase in the Lorentz factor of the jet, causes an increase in the synchrotron emission, while at the same time increasing the gamma-ray opacity due to electron-positron pair production. This scenario also explains the timing of the spectral index change in the radio emission.

The change in the spectral index at radio wavelengths can also be explained by a change in the opacity at radio wavelengths. Since the spectral index moves to negative values after this change, this would mean a decrease in opacity at the lower frequencies. While, in contrast, an opacity increase, would result in a flattening of the spectral index \citep{Kellermann1988}.

We also analyzed the kinematics of the feature B and found that it has a superluminal proper motion of $(3.70 \pm 0.35)c$, which implies that a gamma-ray emission zone within 3C~279 is changing its location with time.

The suggested identification of the gamma-ray emission zone with a region inside the jet in 3C~279 will foster a better understanding of the emission stratification at pc scales. These kind of studies will also help improve future modeling attempts for time-resolved SED analysis, since the changes on the model parameters should also be constrained with the information obtained from the broadband light curves and changes observed at pc-scale.

%%%%%%%%%%%%%%%%%%%%%%%%%%%%%%%%%%%%%%
%%%%%%%%%%%%% ACKNOWLEDGMENTS %%%%%%%%%%%%
%%%%%%%%%%%%%%%%%%%%%%%%%%%%%%%%%%%%%%

\begin{acknowledgements}
This work was supported by CONACyT research grant 280789 (M\'exico). This research has made use of data from the OVRO 40-m monitoring program \citep{Richards2011} which is supported in part by NASA grants NNX08AW31G, NNX11A043G, and NNX14AQ89G and NSF grants AST-0808050 and AST-1109911.
This research has made use of data from the MOJAVE database that is maintained by the MOJAVE team \citep{Lister2009}.
This study makes use of 43 GHz VLBA data from the VLBA-BU Blazar Monitoring Program (VLBA-BU-BLAZAR;
http://www.bu.edu/blazars/VLBAproject.html), funded by NASA through the Fermi Guest Investigator Program. 
The VLBA is an instrument of the Long Baseline Observatory. The Long Baseline Observatory is a facility of the National Science Foundation operated by Associated Universities, Inc. 
\end{acknowledgements}

\bibliographystyle{aa} 

\bibliography{3C279_refs} 

\begin{thebibliography}{40}
\expandafter\ifx\csname natexlab\endcsname\relax\def\natexlab#1{#1}\fi

\bibitem[{{Abdo} {et~al.}(2010){Abdo}, {Ackermann}, {Ajello}, {Allafort},
  {Antolini}, {Atwood}, {Axelsson}, {Baldini}, {Ballet}, {Barbiellini}, \&
  et~al.}]{Abdo2010c}
{Abdo}, A.~A., {Ackermann}, M., {Ajello}, M., {et~al.} 2010, ApJS, 188, 405

\bibitem[{Abdo {et~al.}(2009)Abdo, Ackermann, Ajello, Atwood, Axelsson,
  Baldini, \& et~al.}]{Abdo2009}
Abdo, A.~A., Ackermann, M., Ajello, M., {et~al.} 2009, ApJS, 183, 46

\bibitem[{{Acero} {et~al.}(2015){Acero}, {Ackermann}, {Ajello}, {Albert},
  {Atwood}, {Axelsson}, {Baldini}, {Ballet}, {Barbiellini}, {Bastieri},
  {Belfiore}, {Bellazzini}, {Bissaldi}, {Blandford}, {Bloom}, {Bogart},
  {Bonino}, {Bottacini}, {Bregeon}, {Britto}, {Bruel}, {Buehler}, {Burnett},
  {Buson}, {Caliandro}, {Cameron}, {Caputo}, {Caragiulo}, {Caraveo},
  {Casandjian}, {Cavazzuti}, {Charles}, {Chaves}, {Chekhtman}, {Cheung},
  {Chiang}, {Chiaro}, {Ciprini}, {Claus}, {Cohen-Tanugi}, {Cominsky}, {Conrad},
  {Cutini}, {D'Ammando}, {de Angelis}, {DeKlotz}, {de Palma}, {Desiante},
  {Digel}, {Di Venere}, {Drell}, {Dubois}, {Dumora}, {Favuzzi}, {Fegan},
  {Ferrara}, {Finke}, {Franckowiak}, {Fukazawa}, {Funk}, {Fusco}, {Gargano},
  {Gasparrini}, {Giebels}, {Giglietto}, {Giommi}, {Giordano}, {Giroletti},
  {Glanzman}, {Godfrey}, {Grenier}, {Grondin}, {Grove}, {Guillemot}, {Guiriec},
  {Hadasch}, {Harding}, {Hays}, {Hewitt}, {Hill}, {Horan}, {Iafrate}, {Jogler},
  {J{\'o}hannesson}, {Johnson}, {Johnson}, {Johnson}, {Johnson}, {Kamae},
  {Kataoka}, {Katsuta}, {Kuss}, {La Mura}, {Landriu}, {Larsson}, {Latronico},
  {Lemoine-Goumard}, {Li}, {Li}, {Longo}, {Loparco}, {Lott}, {Lovellette},
  {Lubrano}, {Madejski}, {Massaro}, {Mayer}, {Mazziotta}, {McEnery},
  {Michelson}, {Mirabal}, {Mizuno}, {Moiseev}, {Mongelli}, {Monzani},
  {Morselli}, {Moskalenko}, {Murgia}, {Nuss}, {Ohno}, {Ohsugi}, {Omodei},
  {Orienti}, {Orlando}, {Ormes}, {Paneque}, {Panetta}, {Perkins},
  {Pesce-Rollins}, {Piron}, {Pivato}, {Porter}, {Racusin}, {Rando}, {Razzano},
  {Razzaque}, {Reimer}, {Reimer}, {Reposeur}, {Rochester}, {Romani},
  {Salvetti}, {S{\'a}nchez-Conde}, {Saz Parkinson}, {Schulz}, {Siskind},
  {Smith}, {Spada}, {Spandre}, {Spinelli}, {Stephens}, {Strong}, {Suson},
  {Takahashi}, {Takahashi}, {Tanaka}, {Thayer}, {Thayer}, {Thompson},
  {Tibaldo}, {Tibolla}, {Torres}, {Torresi}, {Tosti}, {Troja}, {Van Klaveren},
  {Vianello}, {Winer}, {Wood}, {Wood}, {Zimmer}, \& {Fermi-LAT
  Collaboration}}]{Acero2015}
{Acero}, F., {Ackermann}, M., {Ajello}, M., {et~al.} 2015, ApJS, 218, 23

\bibitem[{{Ackermann} {et~al.}(2016){Ackermann}, {Anantua}, {Asano}, {Baldini},
  {Barbiellini}, {Bastieri}, {Becerra Gonzalez}, {Bellazzini}, {Bissaldi}, \&
  {Blandford}}]{Ackermann2016}
{Ackermann}, M., {Anantua}, R., {Asano}, K., {et~al.} 2016, ApJL, 824, L20

\bibitem[{{Aleksi{\'c}} {et~al.}(2011){Aleksi{\'c}}, {Antonelli}, {Antoranz},
  {Backes}, {Barrio}, {Bastieri}, {Becerra Gonz{\'a}lez}, {Bednarek},
  {Berdyugin}, {Berger}, {Bernardini}, {Biland}, {Blanch}, {Bock}, {Boller},
  {Bonnoli}, {Borla Tridon}, {Braun}, {Bretz}, {Ca{\~n}ellas}, {Carmona},
  {Carosi}, {Colin}, {Colombo}, {Contreras}, {Cortina}, {Cossio}, {Covino},
  {Dazzi}, {de Angelis}, {de Cea Del Pozo}, {de Lotto}, {Delgado Mendez},
  {Diago Ortega}, {Doert}, {Dom{\'{\i}}nguez}, {Dominis Prester}, {Dorner},
  {Doro}, {Elsaesser}, {Ferenc}, {Fonseca}, {Font}, {Fruck}, {Garc{\'{\i}}a
  L{\'o}pez}, {Garczarczyk}, {Garrido}, {Giavitto}, {Godinovi{\'c}}, {Hadasch},
  {H{\"a}fner}, {Herrero}, {Hildebrand}, {Hose}, {Hrupec}, {Huber}, {Jogler},
  {Klepser}, {Kr{\"a}henb{\"u}hl}, {Krause}, {La Barbera}, {Lelas}, {Leonardo},
  {Lindfors}, {Lombardi}, {L{\'o}pez}, {Lorenz}, {Majumdar}, {Makariev},
  {Maneva}, {Mankuzhiyil}, {Mannheim}, {Maraschi}, {Mariotti},
  {Mart{\'{\i}}nez}, {Mazin}, {Meucci}, {Miranda}, {Mirzoyan}, {Miyamoto},
  {Mold{\'o}n}, {Moralejo}, {Nieto}, {Nilsson}, {Orito}, {Oya}, {Paoletti},
  {Pardo}, {Paredes}, {Partini}, {Pasanen}, {Pauss}, {Perez-Torres}, {Persic},
  {Peruzzo}, {Pilia}, {Pochon}, {Prada}, {Prada Moroni}, {Prandini}, {Puljak},
  {Reichardt}, {Reinthal}, {Rhode}, {Rib{\'o}}, {Rico}, {R{\"u}gamer},
  {R{\"u}ger}, {Saggion}, {Saito}, {Saito}, {Salvati}, {Satalecka},
  {Scalzotto}, {Scapin}, {Schultz}, {Schweizer}, {Shayduk}, {Shore},
  {Sillanp{\"a}{\"a}}, {Sitarek}, {Sobczynska}, {Spanier}, {Spiro}, {Stamerra},
  {Steinke}, {Storz}, {Strah}, {Suri{\'c}}, {Takalo}, {Tavecchio}, {Temnikov},
  {Terzi{\'c}}, {Tescaro}, {Teshima}, {Thom}, {Tibolla}, {Torres}, {Treves},
  {Vankov}, {Vogler}, {Wagner}, {Weitzel}, {Zabalza}, {Zandanel}, \&
  {Zanin}}]{Aleksic2011}
{Aleksi{\'c}}, J., {Antonelli}, L.~A., {Antoranz}, P., {et~al.} 2011, A\&A,
  530, A4

\bibitem[{{Arshakian} {et~al.}(2010){Arshakian}, {Le{\'o}n-Tavares}, {Lobanov},
  {Chavushyan}, {Shapovalova}, {Burenkov}, \& {Zensus}}]{Arshakian2010}
{Arshakian}, T.~G., {Le{\'o}n-Tavares}, J., {Lobanov}, A.~P., {et~al.} 2010,
  MNRAS, 401, 1231

\bibitem[{{Bonning} {et~al.}(2012){Bonning}, {Urry}, {Bailyn}, {Buxton},
  {Chatterjee}, {Coppi}, {Fossati}, {Isler}, \& {Maraschi}}]{Bonning2012}
{Bonning}, E., {Urry}, C.~M., {Bailyn}, C., {et~al.} 2012, ApJ, 756, 13

\bibitem[{{B{\"o}ttcher} {et~al.}(2013){B{\"o}ttcher}, {Reimer}, {Sweeney}, \&
  {Prakash}}]{Bottcher2013}
{B{\"o}ttcher}, M., {Reimer}, A., {Sweeney}, K., \& {Prakash}, A. 2013, ApJ,
  768, 54

\bibitem[{{Chatterjee} {et~al.}(2008){Chatterjee}, {Jorstad}, {Marscher}, {Oh},
  {McHardy}, {Aller}, {Aller}, {Balonek}, {Miller}, {Ryle}, {Tosti},
  {Kurtanidze}, {Nikolashvili}, {Larionov}, \& {Hagen-Thorn}}]{Chatterjee2008}
{Chatterjee}, R., {Jorstad}, S.~G., {Marscher}, A.~P., {et~al.} 2008, ApJ, 689,
  79

\bibitem[{{Collmar} {et~al.}(2010){Collmar}, {B{\"o}ttcher}, {Krichbaum},
  {Agudo}, {Bottacini}, {Bremer}, {Burwitz}, {Cuccchiara}, {Grupe}, \&
  {Gurwell}}]{Collmar2010}
{Collmar}, W., {B{\"o}ttcher}, M., {Krichbaum}, T.~P., {et~al.} 2010, A\&A,
  522, A66

\bibitem[{{Gurwell} {et~al.}(2007){Gurwell}, {Peck}, {Hostler}, {Darrah}, \&
  {Katz}}]{Gurwell2007}
{Gurwell}, M.~A., {Peck}, A.~B., {Hostler}, S.~R., {Darrah}, M.~R., \& {Katz},
  C.~A. 2007, in Astronomical Society of the Pacific Conference Series, Vol.
  375, From Z-Machines to ALMA: (Sub)Millimeter Spectroscopy of Galaxies, ed.
  A.~J. {Baker}, J.~{Glenn}, A.~I. {Harris}, J.~G. {Mangum}, \& M.~S. {Yun},
  234

\bibitem[{{Hartman} {et~al.}(1992){Hartman}, {Bertsch}, {Fichtel}, {Hunter},
  {Kanbach}, {Kniffen}, \& et~al.}]{Hartman1992}
{Hartman}, R.~C., {Bertsch}, D.~L., {Fichtel}, C.~E., {et~al.} 1992, ApJL, 385,
  L1

\bibitem[{{Hayashida} {et~al.}(2012){Hayashida}, {Madejski}, {Nalewajko},
  {Sikora}, {Wehrle}, {Ogle}, {Collmar}, {Larsson}, {Fukazawa}, {Itoh},
  {Chiang}, {Stawarz}, {Blandford}, {Richards}, {Max-Moerbeck}, {Readhead},
  {Buehler}, {Cavazzuti}, {Ciprini}, {Gehrels}, {Reimer}, {Szostek}, {Tanaka},
  {Tosti}, {Uchiyama}, {Kawabata}, {Kino}, {Sakimoto}, {Sasada}, {Sato},
  {Uemura}, {Yamanaka}, {Greiner}, {Kruehler}, {Rossi}, {Macquart}, {Bock},
  {Villata}, {Raiteri}, {Agudo}, {Aller}, {Aller}, {Arkharov}, {Bach},
  {Ben{\'{\i}}tez}, {Berdyugin}, {Blinov}, {Blumenthal}, {B{\"o}ttcher},
  {Buemi}, {Carosati}, {Chen}, {Di Paola}, {Dolci}, {Efimova}, {Forn{\'e}},
  {G{\'o}mez}, {Gurwell}, {Heidt}, {Hiriart}, {Jordan}, {Jorstad}, {Joshi},
  {Kimeridze}, {Konstantinova}, {Kopatskaya}, {Koptelova}, {Kurtanidze},
  {L{\"a}hteenm{\"a}ki}, {Lamerato}, {Larionov}, {Larionova}, {Larionova},
  {Leto}, {Lindfors}, {Marscher}, {McHardy}, {Molina}, {Morozova},
  {Nikolashvili}, {Nilsson}, {Reinthal}, {Roustazadeh}, {Sakamoto}, {Sigua},
  {Sillanp{\"a}{\"a}}, {Takalo}, {Tammi}, {Taylor}, {Tornikoski}, {Trigilio},
  {Troitsky}, \& {Umana}}]{Hayashida2012}
{Hayashida}, M., {Madejski}, G.~M., {Nalewajko}, K., {et~al.} 2012, \apj, 754,
  114

\bibitem[{{Hayashida} {et~al.}(2015){Hayashida}, {Nalewajko}, {Madejski},
  {Sikora}, {Itoh}, {Ajello}, \& et~al.}]{Hayashida2015}
{Hayashida}, M., {Nalewajko}, K., {Madejski}, G.~M., {et~al.} 2015, ApJ, 807,
  79

\bibitem[{{Janiak} {et~al.}(2012){Janiak}, {Sikora}, {Nalewajko}, {Moderski},
  \& {Madejski}}]{Janiak2012}
{Janiak}, M., {Sikora}, M., {Nalewajko}, K., {Moderski}, R., \& {Madejski},
  G.~M. 2012, ApJ, 760, 129

\bibitem[{{Jorstad} {et~al.}(2017){Jorstad}, {Marscher}, {Morozova},
  {Troitsky}, {Agudo}, {Casadio}, {Foord}, {G{\'o}mez}, {MacDonald}, {Molina},
  {L{\"a}hteenm{\"a}ki}, {Tammi}, \& {Tornikoski}}]{Jorstad2017}
{Jorstad}, S.~G., {Marscher}, A.~P., {Morozova}, D.~A., {et~al.} 2017, ApJ,
  846, 98

\bibitem[{{Kellermann} \& {Owen}(1988)}]{Kellermann1988}
{Kellermann}, K.~I. \& {Owen}, F.~N. 1988, {Radio galaxies and quasars}, ed.
  K.~I. {Kellermann} \& G.~L. {Verschuur}, 563--602

\bibitem[{{Larionov} {et~al.}(2008){Larionov}, {Jorstad}, {Marscher},
  {Raiteri}, {Villata}, {Agudo}, \& et~al.}]{Larionov2008}
{Larionov}, V.~M., {Jorstad}, S.~G., {Marscher}, A.~P., {et~al.} 2008, A\&A,
  492, 389

\bibitem[{{Le{\'o}n-Tavares} {et~al.}(2010){Le{\'o}n-Tavares}, {Lobanov},
  {Chavushyan}, {Arshakian}, {Doroshenko}, {Sergeev}, {Efimov}, \&
  {Nazarov}}]{Leon-Tavares2010}
{Le{\'o}n-Tavares}, J., {Lobanov}, A.~P., {Chavushyan}, V.~H., {et~al.} 2010,
  ApJ, 715, 355

\bibitem[{{Lindfors} {et~al.}(2006){Lindfors}, {T{\"u}rler}, {Valtaoja},
  {Aller}, {Aller}, {Mazin}, {Raiteri}, {Stevens}, {Tornikoski}, {Tosti}, \&
  {Villata}}]{Lindfors2006}
{Lindfors}, E.~J., {T{\"u}rler}, M., {Valtaoja}, E., {et~al.} 2006, A\&A, 456,
  895

\bibitem[{{Lister} {et~al.}(2009){Lister}, {Aller}, {Aller}, {Cohen}, {Homan},
  {Kadler}, {Kellermann}, {Kovalev}, {Ros}, {Savolainen}, {Zensus}, \&
  {Vermeulen}}]{Lister2009}
{Lister}, M.~L., {Aller}, H.~D., {Aller}, M.~F., {et~al.} 2009, AJ, 137, 3718

\bibitem[{{Lister} {et~al.}(2013){Lister}, {Aller}, {Aller}, {Homan},
  {Kellermann}, {Kovalev}, {Pushkarev}, {Richards}, {Ros}, \&
  {Savolainen}}]{Lister2013}
{Lister}, M.~L., {Aller}, M.~F., {Aller}, H.~D., {et~al.} 2013, AJ, 146, 120

\bibitem[{{Lister} {et~al.}(2016){Lister}, {Aller}, {Aller}, {Homan},
  {Kellermann}, {Kovalev}, {Pushkarev}, {Richards}, {Ros}, \&
  {Savolainen}}]{Lister2016}
{Lister}, M.~L., {Aller}, M.~F., {Aller}, H.~D., {et~al.} 2016, AJ, 152, 12

\bibitem[{{Marscher} {et~al.}(2008){Marscher}, {Jorstad}, {D'Arcangelo},
  {Smith}, {Williams}, {Larionov}, \& et~al.}]{Marscher2008}
{Marscher}, A.~P., {Jorstad}, S.~G., {D'Arcangelo}, F.~D., {et~al.} 2008,
  Nature, 452, 966

\bibitem[{{McMullin} {et~al.}(2007){McMullin}, {Waters}, {Schiebel}, {Young},
  \& {Golap}}]{McMullin2007}
{McMullin}, J.~P., {Waters}, B., {Schiebel}, D., {Young}, W., \& {Golap}, K.
  2007, in Astronomical Society of the Pacific Conference Series, Vol. 376,
  Astronomical Data Analysis Software and Systems XVI, ed. R.~A. {Shaw},
  F.~{Hill}, \& D.~J. {Bell}, 127

\bibitem[{{Nolan} {et~al.}(2012){Nolan}, {Abdo}, {Ackermann}, {Ajello},
  {Allafort}, {Antolini}, {Atwood}, {Axelsson}, {Baldini}, {Ballet}, \&
  et~al.}]{Nolan2012}
{Nolan}, P.~L., {Abdo}, A.~A., {Ackermann}, M., {et~al.} 2012, ApJS, 199, 31

\bibitem[{{Pati{\~n}o-{\'A}lvarez} {et~al.}(2013){Pati{\~n}o-{\'A}lvarez},
  {Chavushyan}, {Le{\'o}n-Tavares}, {Vald{\'e}s}, {Carrami{\~n}ana},
  {Carrasco}, \& {Torrealba}}]{Patino-Alvarez2013a}
{Pati{\~n}o-{\'A}lvarez}, V., {Chavushyan}, V., {Le{\'o}n-Tavares}, J.,
  {et~al.} 2013, ArXiv e-prints [\eprint[arXiv]{1303.1893}]

\bibitem[{{Pati{\~n}o-{\'A}lvarez} {et~al.}(2018){Pati{\~n}o-{\'A}lvarez},
  {Fernandes}, {Chavushyan}, {L{\'o}pez-Rodr{\'{\i}}guez}, {Le{\'o}n-Tavares},
  {Schlegel}, {Carrasco}, {Vald{\'e}s}, \&
  {Carrami{\~n}ana}}]{Patino-Alvarez2018}
{Pati{\~n}o-{\'A}lvarez}, V.~M., {Fernandes}, S., {Chavushyan}, V., {et~al.}
  2018, MNRAS, 479, 2037

\bibitem[{Pati{\~n}o-{\'A}lvarez {et~al.}(2017)Pati{\~n}o-{\'A}lvarez,
  Fernandes, Chavushyan, L{\'o}pez-Rodr{\'\i}guez, Le{\'o}n-Tavares, Schlegel,
  Carrasco, Vald{\'e}s, \& Carrami{\~n}ana}]{Patino-Alvarez2017}
Pati{\~n}o-{\'A}lvarez, V.~M., Fernandes, S., Chavushyan, V., {et~al.} 2017,
  FrASS, 4, 47

\bibitem[{{Pauliny-Toth} \& {Kellermann}(1966)}]{Pauliny-Toth1966}
{Pauliny-Toth}, I.~I.~K. \& {Kellermann}, K.~I. 1966, ApJ, 146, 634

\bibitem[{{Pittori} {et~al.}(2018){Pittori}, {Lucarelli}, {Verrecchia},
  {Raiteri}, {Villata}, {Vittorini}, {Tavani}, {Puccetti}, {Perri}, \&
  {Donnarumma}}]{Pittori2018}
{Pittori}, C., {Lucarelli}, F., {Verrecchia}, F., {et~al.} 2018, ApJ, 856, 99

\bibitem[{{Rani} {et~al.}(2018){Rani}, {Jorstad}, {Marscher}, {Agudo},
  {Sokolovsky}, {Larionov}, {Smith}, {Mosunova}, {Borman}, {Grishina},
  {Kopatskaya}, {Mokrushina}, {Morozova}, {Savchenko}, {Troitskaya},
  {Troitsky}, {Thum}, {Molina}, \& {Casadio}}]{Rani2018}
{Rani}, B., {Jorstad}, S.~G., {Marscher}, A.~P., {et~al.} 2018, ApJ, 858, 80

\bibitem[{{Richards} {et~al.}(2011){Richards}, {Max-Moerbeck}, {Pavlidou},
  {King}, {Pearson}, {Readhead}, {Reeves}, {Shepherd}, {Stevenson},
  {Weintraub}, {Fuhrmann}, {Angelakis}, {Zensus}, {Healey}, {Romani}, {Shaw},
  {Grainge}, {Birkinshaw}, {Lancaster}, {Worrall}, {Taylor}, {Cotter}, \&
  {Bustos}}]{Richards2011}
{Richards}, J.~L., {Max-Moerbeck}, W., {Pavlidou}, V., {et~al.} 2011, ApJS,
  194, 29

\bibitem[{{Rodr{\'{\i}}guez} {et~al.}(2017){Rodr{\'{\i}}guez}, {Dzib},
  {Loinard}, {Zapata}, {G{\'o}mez}, {Menten}, \& {Lizano}}]{Rodriguez2017}
{Rodr{\'{\i}}guez}, L.~F., {Dzib}, S.~A., {Loinard}, L., {et~al.} 2017, ApJ,
  834, 140

\bibitem[{{Schinzel} {et~al.}(2012){Schinzel}, {Lobanov}, {Taylor}, {Jorstad},
  {Marscher}, \& {Zensus}}]{Schinzel2012}
{Schinzel}, F.~K., {Lobanov}, A.~P., {Taylor}, G.~B., {et~al.} 2012, A\&A, 537,
  A70

\bibitem[{{Smith} {et~al.}(2009){Smith}, {Montiel}, {Rightley}, {Turner},
  {Schmidt}, \& {Jannuzi}}]{Smith2009}
{Smith}, P.~S., {Montiel}, E., {Rightley}, S., {et~al.} 2009, ArXiv e-prints
  [\eprint[arXiv]{0912.3621}]

\bibitem[{{The Fermi-LAT collaboration}(2019)}]{4FGL}
{The Fermi-LAT collaboration}. 2019, arXiv e-prints, arXiv:1902.10045

\bibitem[{{Torrealba} {et~al.}(2012){Torrealba}, {Chavushyan},
  {Cruz-Gonz{\'a}lez}, {Arshakian}, {Bertone}, \&
  {Rosa-Gonz{\'a}lez}}]{Torrealba2012}
{Torrealba}, J., {Chavushyan}, V., {Cruz-Gonz{\'a}lez}, I., {et~al.} 2012,
  RMxAA, 48, 9

\bibitem[{{Urry} \& {Padovani}(1995)}]{Urry1995}
{Urry}, C.~M. \& {Padovani}, P. 1995, PASP, 107, 803

\bibitem[{Vaughan {et~al.}(2003)Vaughan, Edelson, Warwick, \&
  Uttley}]{Vaughan2003}
Vaughan, S., Edelson, R., Warwick, R.~S., \& Uttley, P. 2003, MNRAS, 345, 1271

\end{thebibliography}

\clearpage

%%%%%%%%%%%%%%%%%%%%%%%%%%%%%%%%%%%%%%
%%%%%%%%%%%%%%%% APPENDIX %%%%%%%%%%%%%%%%
%%%%%%%%%%%%%%%%%%%%%%%%%%%%%%%%%%%%%%

\end{document}